# Understanding the Needs of Nonhuman Stakeholders in Design Process: An Overview of and Reflection on Methods

**Berre Su Yanlıç, Arçelik Research Center for Creative Industries, Koç University**

**Aykut Coşkun, Arçelik Research Center for Creative Industries, Koç University**

## ABSTRACT

Design practice traditionally focused on human concerns, either overseeing the various effects of climate issues on nonhuman stakeholders or considering them as resources to address these problems. The climate crisis's urgency demands a design shift towards sustainability and inclusivity. This shift was happening through an emerging theme in design, More-Than-Human (MTH), which expands the notion of the user to animals, things, nature, and microbes. Such an expansion creates a requirement for designers to consider nonhuman perspectives during the design process. This paper investigates the methods used in MTH Design studies to explore and synthesize the perspectives of nonhuman users. Reviewing 30 papers, it highlights a predominant focus on animals and things over plants and microbes in MTH studies, along with a scarcity of synthesis methods. It identifies the necessity of tools that represent nonhumans with their relationships within larger ecosystems, and calls for increased attention to plants and microbes, emphasizing their vital role in sustainable environments and urging researchers to develop methods for understanding these species. By highlighting method strengths and weaknesses, it aims to guide designers and design researchers who plan to work with nonhuman users in selecting appropriate methods.

## Introduction

Climate change is a serious threat to our planet, and immediate action is needed to transition to a more sustainable future. As we deal with the effects of environmental degradation, such as the continuous rise of carbon dioxide levels and increased frequency of heat waves, droughts, and floods (United Nations, 2023), the significance of implementing sustainable practices becomes evident.

Design could contribute to sustainable futures as a discipline with diverse tools and methods to address wicked problems. Previous studies on sustainable design explored various facets of such a contribution, concentrating on environmental, social, and economic sustainability (Scuri et al., 2022). While earlier sustainable design approaches focused on product-level solutions (e.g. eco-design, green design, biomimicry), in time, this focus has shifted to system-level solutions (e.g. Product Service Systems, Design for Systems of Innovations and Transitions) (Ceschin & Gaziulusoy, 2016), and more recently, the field embraced the notion of More-than-human (MTH) Design.

MTH Design's approach to sustainability aims to make the well-being of diverse stakeholders, including animals, plants, and ecosystems, central to sustainable design solutions. This approach is based on the notion that humans are not the only species influenced by the negative consequences of environmental problems, and that humans and nonhumans are part of a bigger network (Haraway, 2013; Forlano, 2016), co-existing and entangled (Häggström, 2019; Westerlaken & Gualeni, 2016). For instance, extreme heat and droughts have a direct or indirect negative impact on animal health and well-being (Lacetera, 2019). Terrestrial biodiversity is also being lost because of altered land use, such as the expansion of farmlands (Malhi et al., 2020). Thus, approaching these issues from a human-centered perspective imposes limitations. To build more "inclusive" sustainable futures (Markard et al., 2012), the perspectives and needs of nonhuman stakeholders should be considered alongside those of humans, and relationships between humans and nonhumans should be considered in design (Forlano, 2016; J. Liu et al., 2018; Tomitsch et al., 2021).

The introduction of MTH and Post-humanistic values to design literature (Cox et al., 2020; Behzad et al., 2022; Oogjes & Wakkary, 2022) expanded the definition of "user" from its initial conception, which represents humans as the primary users of products, services, and systems, towards a concept including both humans and nonhumans as users (Coskun et al., 2022; Oogjes & Wakkary, 2022). These nonhumans include animals (Smith et al., 2017; Phillips & Kau, 2019; Biggs et al., 2021; Heitlinger et al., 2021; Wolff et al., 2021), things[1] (Chang et al., 2017; Giaccardi et al., 2020; Reddy, Kocaballi, et al., 2021), ecological systems (Pettersen et al., 2018; S.-Y. (Cyn) Liu et al., 2018; Biggs & Desjardins, 2020; Pollastri et al., 2021a; Wolff et al., 2021) and microorganisms (Chen et al., 2021).

Although considering nonhuman perspectives is crucial for a sustainable future, including their perspectives in the design process and shaping the design outcomes in alignment with their needs is challenging for design researchers and practitioners. This is due to humans' tendency to see nonhumans through human practices (Harrison & Hall, 2010), and undeniable biological differences between them, such as being able to speak (Hastrup, 2015). Hence, the effectiveness of existing human-centered design thinking methods (elicitation techniques like interviews, surveys, focus groups, synthesis techniques like persona, empathy maps, and scenarios) in exploring the perspectives of nonhuman stakeholders is questionable.

Previously, design researchers employed various methods to mitigate this challenge. For instance, they used noticing (Oogjes & Wakkary, 2022) and contact zone (Prost et al., 2021) to facilitate encounters with nonhumans and reflections on these encounters, or animal persona (Hirskyj-Douglas et al., 2017) to represent animals in the design process. While numerous techniques have been proposed to understand nonhuman perspectives in the design thinking process, the dominance of human-centered methods persists, and the diversity of methods focused on MTH perspectives remains limited. Additionally, the lack of an overview of these methods makes it difficult to identify their benefits and drawbacks, thus preventing the design research community from selecting the appropriate methods.

[1] We use the concepts of "thing" and "object" as synonyms.

In this paper, we aim to address this gap by reviewing 30 papers that utilize an MTH Design approach, looking closely into the methods used in the early stages of the design process to elicit and synthesize the requirements of nonhuman users. Our review showed that methods to synthesize the nonhuman perspectives are fewer than methods to explore these perspectives and that studies focusing on plants and microbes represent a minority compared to studies focusing on animals and things. Our findings also revealed a transition from methods addressing a single nonhuman to those addressing multiple nonhumans. Besides these, we contribute to the literature by categorizing the methods based on their objectives (i.e. reflection, imagining, and data gathering) and by presenting their strengths and weaknesses. Thus, this paper could be a valuable source for design researchers and practitioners interested in MTH Design in selecting the appropriate method for exploring and synthesizing nonhuman users' needs and perspectives.

## Related Work

### More-Than-Human Design and Sustainability

MTH Design utilizes diverse theories and practices connected to different fields, including posthuman (Forlano, 2017), new materialism (Braidotti, 2013), feminism (Weheliye, 2014), actor-network (Latour, 1996), and object-oriented ontology (Harman, 2017). It argues that humans are not the only users of design outcomes (Coulton & Lindley, 2019), and thus designers should account for nonhuman stakeholders (French et al., 2020; Hirskyj-Douglas et al., 2017), promote ecological sustainability and interspecies well-being (Kirk et al., 2019; Porter, 2019), and design for co-habitation (Tironi et al., 2023) rather than only focusing on humans as users.

As addressing sustainability solely through a human-focused lens, without acknowledging their interdependence with nonhumans, is insufficient for achieving a sustainable future (Rupprecht et al., 2020), MTH perspectives have been explored and found resonance in sustainable design (Coskun et al., 2022). Examples include exploring how to include MTH perspectives in urban design (Clarke et al., 2018, 2019; Metzger & Lindblad, 2020), MTH spatial planning practices (Metzger, 2014, 2019; Metzger & Lindblad, 2020), MTH media architecture with a focus on self-owning autonomous forests (Sheikh et al., 2021), translating plants' experiences into the human language (Steiner et al., 2017), using design to arouse curiosity about nature (Portocarrero et al., 2015), and foster empathy towards microbes (Chen et al., 2021). Commercial examples include products such as PlantWave (*PlantWave*, 2020) and artworks such as Phonopholium to foster interspecies interactions (*Scencosme – Phonopholium*, 2018).

Design thinking is a user-centered process that prioritizes empathizing with users of a product, service, or system (Liedtka Jeanne & Ogilvie Tim, 2011; Wolniak, 2017). While the involvement of users should be considered for each stage, exploring and synthesizing user needs are critical for understanding the nature of a design challenge, uncovering hidden needs and opportunities, and defining the right problem. Looking at the studies focusing on exploring and synthesizing MTH perspectives, it seems that designers and researchers utilize either adapted methods or develop new ones. The former category includes, for instance, animal persona, which is adapted from persona to represent animal stakeholders, and noticing, which is based on autoethnography to foster reflection by engaging with nonhumans in the field. The latter category includes Animal Diplomacy Bureau's Bird Games which let participants experience the world from a bird's perspective, Umwelt-Sketch which aims to visualize how various species perceive and interact with their environment, and Breathing-With which fosters awareness of interconnected relationships between humans and nonhumans through breathing.

Understanding these methods and their use in the design process is crucial for incorporating nonhuman perspectives effectively. This ensures that the resulting ideas, artifacts, or systems from the design process are adequately tailored to meet the needs of nonhumans. Furthermore, examining the use of these methods closely is essential for recognizing their advantages and disadvantages and identifying methodological gaps in the literature. Despite the multitude of techniques used to explore and synthesize nonhuman perspectives in the design process, the literature lacks a study analyzing and providing an overview of these techniques. In this paper, we address this gap through an analysis of 30 papers in the MTH Design field.

## Methodology

To find the relevant papers for the review, we used Web of Science as a general-purpose database, the Design Research Society Digital Library, and ACM Digital Library as domain-specific libraries. We searched these venues using the Anywhere function for the following conditions:

1. "More-Than-Human" + Perspective
2. "More-Than-Human" + Understanding
3. "More-Than-Human" + Empathy

Then, we searched again with the "More-Than-Human" condition to identify any missing papers from the previous search. This yielded a total of 2008 papers.

After this collection, we excluded the overlapping papers and filtered the remaining papers according to the following criteria:

1. At least one method used/described in the study should be used for the exploration or synthesis of nonhuman needs.

2. The methods should be used in a design study.
3. The paper should be in English.

Furthermore, we filtered Extended Abstracts and Demonstration papers. This resulted in the selection of 30 papers. We performed our analysis with this set of papers, by asking the following questions:

1. Which nonhuman users were included in the study?
2. How were these users involved? Which data collection method(s) were used?
3. What was the purpose of the methods in the study?

As a final step, we created a sub-group out of the papers in which the authors reflected on the methods they used/described. This sub-group included 17 papers, which we analyzed according to the advantages and disadvantages of the methods (based on the reflections of the authors of these papers).

It should be noted that our aim was not to conduct a systematic review but rather to cohesively present experiences and lessons learned from researchers working with nonhumans in the field. Plus, as our focus is on sustainability, we initially planned to include papers focusing on living entities (animals, plants, microbes, inanimate natural), excluding papers focusing on things. However, thinking that these methods could be adapted for living entities in the future and that there might be important lessons for their adaption we decided to include them in our analysis.

## Results

In this section, we first present a broader picture of methods used in the early stages of the design process in the MTH field. Then, we dive into the specifics of these methods, including the contexts in which they are applied, as well as their benefits, challenges, and author recommendations, according to the authors' reflections.

### Overview of the MTH Design Research Methods

Figure 1 presents methods used in the exploration and synthesis stages according to targeted nonhumans (e.g. animals, plants) and activities designers engaged in (e.g. reflection, data gathering, imagining).

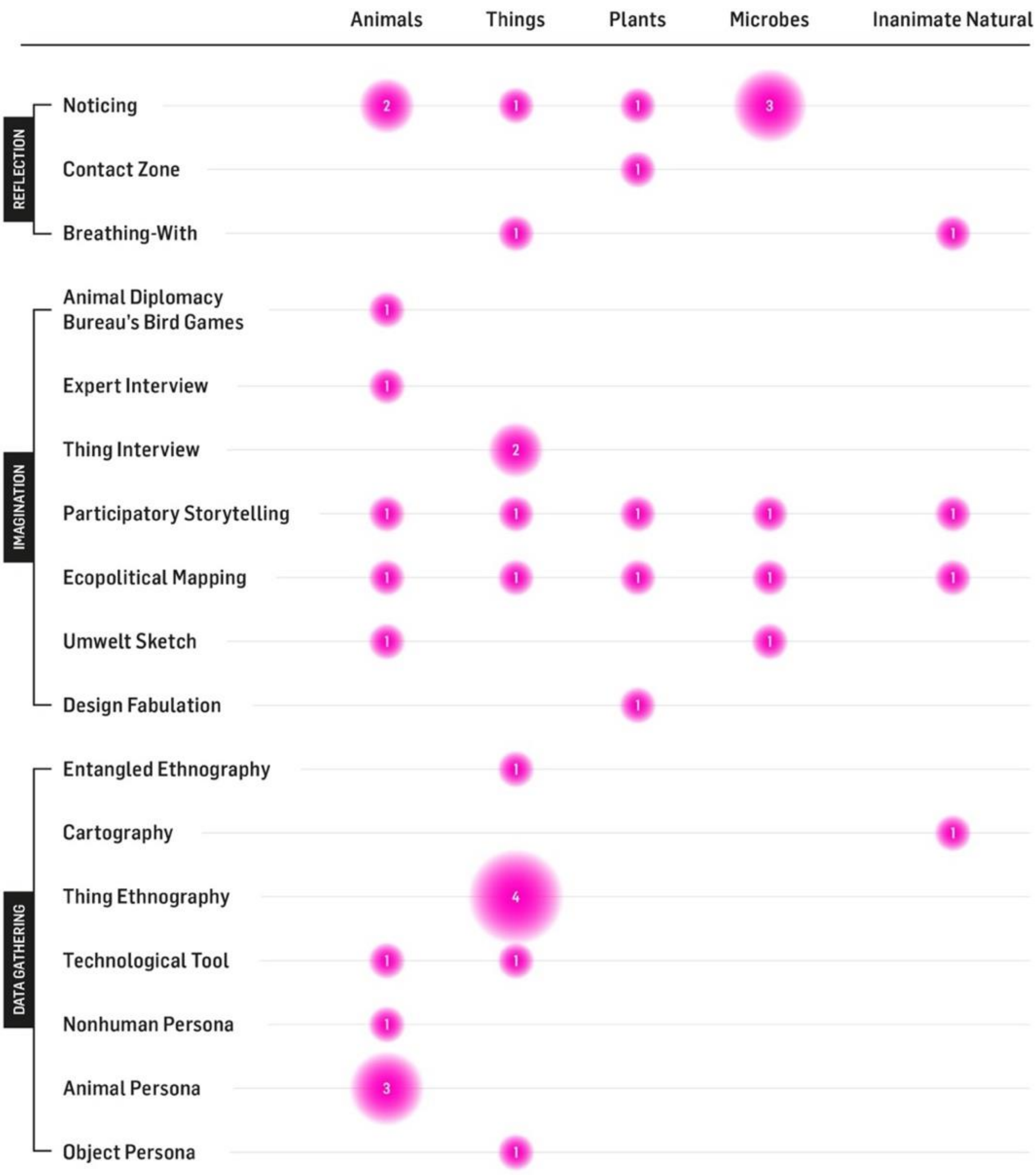

*Figure 1: Mapping of methods used in MTH literature for exploration and synthesis.*

As seen in Fig. 1, when we examine the exploration and synthesis methods used in the literature, we currently observe that these methods are predominantly used for animals and things. For instance, the *Animal Diplomacy Bureau's bird games* (Phillips & Kau, 2019), and *expert interview* (French et al., 2020) have only been used for animals so far. *Thing interview* (Chang et al., 2017; Reddy, Kocaballi, et al., 2021; Reddy, Nicenboim, et al., 2021), *thing ethnography* (Cila et al., 2015; Giaccardi et al., 2016, 2020; Macruz et al., 2023), and *entangled ethnography* (Murray-Rust Katerina Gorkovenko DMurray-Rust et al., 2019) have been used exclusively for things. Similarly, *technological tool*s are used only for things (Cho et al., 2023) and animals (Forkosh, 2021). Additionally, in synthesis methods, *personas* have been used only for animals and things.

Apart from these studies, there are also methods used with plants, microbes, and inanimate natural (air, sea, soil, etc.). These include *contact zone* (Prost et al., 2021) and *design fabulation* (Søndergaard, 2023) with plants, *cartography* (Pollastri et al., 2021) with inanimate natural. *Umwelt-sketch* (Benöhr et al., 2022; De Roo & Ganzevles, 2023) is used with animals and microbes, and *ecopolitical mapping* (Benöhr et al., 2022) is used with animals, plants, inanimate natural, microbiomes, and things.

When we examine these methods according to years, we observe a shift from focusing solely on one nonhuman to exploring or representing multiple nonhumans. For example, *personas* start with *animal persona* (Frawley et al., 2014; Hirskyj-Douglas et al., 2017) and *object persona* (Cila & Giaccardi, 2015) and later evolve into *nonhuman persona* (Tomitsch et al., 2021). Likewise, exploration techniques begin with *thing ethnography* (Cila et al., 2015; Giaccardi et al., 2016, 2020), followed by *noticing* focusing on fungi (J. Liu et al., 2018), animals (Biggs et al., 2021; Livio & Devendorf, 2022), and things (Oogjes & Wakkary, 2022) separately. More recently, we have seen *noticing* focusing on plants and microbiomes together (Rosén et al., 2022), *umwelt-sketch* focusing on animals and microbes (De Roo & Ganzevles, 2023), and *participatory storytelling* focusing on animals, plants, inanimate natural, microbiomes, and things (Talgorn & Ullerup, 2023).

Finally, when comparing exploration and synthesis methods (Fig. 1), we see that the number and diversity of synthesis methods are much lower compared to exploration methods. Out of the 17 methods (from 30 papers) examined in our overview, 14 are utilized for exploration, each distinct from the other. On the contrary, only 3 methods are used for synthesis, all of which are adaptations of the persona method.

## Insight on Methods

Our analysis revealed that the methods used in MTH Design studies can be categorized according to their intended purposes: data gathering, imagination, and reflection.

### *Data Gathering*

Methods whose purpose is data gathering primarily aim to acquire information about nonhumans through various means (e.g. cameras, experts, proxies). *Thing ethnography* and various *persona* methods including *animal persona, object persona*, and *nonhuman persona* are grouped under this purpose.

*Thing ethnography* (Cila et al., 2015) aims to collect information regarding the utilization, mobility, and interconnections among things & humans, and things & things. This enables the observation and analysis of the behaviors and interactions associated with objects within specific contexts. This method involves collecting visual and aural data about daily objects (e.g. kettles, cups, and fridges) with autographs, and interviewing their users about their daily practices. According to the authors, *Thing ethnography* helps to understand nonhuman perspectives and worlds by revealing their patterns (of use, movement, and time), their role in human practices, and analyzing them in specific socio-cultural settings.

*Persona* aims to synthesize the information gathered during exploration to make it usable in the design process. In the MTH context, personas facilitate the representation of nonhumans, stakeholders that cannot articulate their perspectives, within the design team. Via *personas*, nonhumans are represented as individuals with their own stories, wants, and needs in the design process in a specific context. *Personas* in the literature include *nonhuman persona* (Tomitsch et al., 2021), *animal persona* (Frawley et al., 2014; Hirskyj-Douglas et al., 2017; Bain & Tomitsch, 2022;), and *object persona* (Cila & Giaccardi, 2015).

To create these *personas*, designers relied on working with experts such as amphibian biologists to get insights about the behaviors of frogs (Bain & Tomitsch, 2022), interviews with proxies such as farmers about characteristics in chickens (Frawley et al., 2014), on-site observation of nonhumans such as collecting frog calls with herpetology experts (Bain & Tomitsch, 2022), reviewing related literature including articles and reports to collect existing information about nonhumans (Tomitsch et al., 2021), and visual data collection about "behavioral patterns, temporal routines, and spatial movements of objects and their users" (Cila & Giaccardi, 2015). *Personas* commonly include demographic information and inner life aspects such as attitudes, motivations, needs, wants, fears, etc. Additionally, in some cases, *personas* represent nonhuman users' interaction with technology (Hirskyj-Douglas et al., 2017), distribution (Bain & Tomitsch, 2022), and social relations (Cila & Giaccardi, 2015).

According to reflections from all authors using the *persona*, this method helps focus on the end-user requirements of the nonhumans, create a shared understanding of the nonhumans among the design team, assess the product from the perspectives of the nonhumans, and identify different relations between nonhumans and humans. Furthermore, Frawley et al. stated that *persona* helped them challenge dominant practices of the human world where nonhumans are often made invisible, and it created a safe space for the design team to articulate understandings of nonhumans.

The researchers using this method advise to avoid misuse of *persona* as it can lead to solutions detrimental to nonhumans when used to justify the decisions of humans (Tomitsch et al., 2021; Bain & Tomitsch, 2022). They highlight that it is eventually limited to the human mindset and there may be some biases (Frawley et al., 2014; Tomitsch et al., 2021; Bain & Tomitsch, 2022). Furthermore, there is a trade-off between actual representation and generalized representation, and some individual stories may get lost along the way (Hirskyj-Douglas et al., 2017).

### *Imagining*

Methods aimed at fostering imagination primarily focus on integrating data collected about nonhuman stakeholders with imagination to explore nonhuman perspectives. In these methods, data may come from an expert (e.g. expert interview) or a technological tool (e.g. a camera). Under this category, we include six different methods: *narratives, design fabulation, participatory storytelling, thing interview, umwelt-sketch,* and *Animal Diplomacy Bureau's bird games.*

*Narratives* (Turner & Morrison, 2020), *design fabulation* (Søndergaard, 2023), and *participatory storytelling* (Talgorn & Ullerup, 2023) aim to create stories about nonhumans. In these methods, data is collected through various techniques and sources, including memories from proxies (i.e. dogs' walkers) with questionnaires (on characteristics of the walkers and dogs, walking habits, and "negotiation between the walker and their dogs") (Turner & Morrison, 2020), personal knowledge of the participants (no additional data collection was mentioned) (Talgorn & Ullerup, 2023) and visual data from the microscope (Søndergaard, 2023).

Turner and Marrison put together responses from the dog walkers in a loose form and corroborated with the dog walkers, using the narrative inquiry method. Talgorn and Ullerup designed a story template in which participants answered some questions about the character's motivations, surrounding environment, obstacles the character faces, the outcome of the story, and the changes the character went through. Søndergaard recorded 6 scenes in the microscope, added them together, and used 6 different human voices to "advocate on behalf of the mosses". While the final story is in the form of text in *narratives* and *participatory storytelling* including both humans and nonhumans, the product of *design fabulation* is a video including moss.

According to the reflections from all authors using these methods, they help understand the nonhuman world and their perspectives by seeing "the world from the character's perspective", discovering that they also have stories to tell, learning about their characteristics, and treating them as individuals. In addition, *narratives* help decenter humans in design and discover tensions. *Participatory storytelling* helps increase awareness about the nonhuman world, offers a "non-confrontational" way to engage with the nonhuman world as it uses metaphors, fosters "openness to unknown" and facilitates "overcoming prejudices". Furthermore, *design fabulation* challenges dominant values on nonhumans (such as seeing the microscope merely as a scientific tool) and encourages care relationships.

Søndergaard emphasizes that the *fabulations* should refrain from "oversimplifying and decontextualizing relations with nonhumans". Rather they should be "embedded and embodied, based on specific values, meanings, and priorities". Furthermore, she highlights that simply noticing and perceiving is not enough to understand the nonhuman perspective; imagination is also necessary. However, there should be a balance between them (Søndergaard, 2023; Talgorn & Ullerup, 2023).

*Thing interview* (Chang et al., 2017) is a "constructive and speculative" approach that aims to create empathy with things from a first-person point of view adapting an originally human-centered method, the interview. In this method, the interviewee role-plays as a nonhuman (scooter) and attempts to respond to the interviewer's questions from the nonhuman's perspective. In doing so, they utilize the information gathered about the nonhuman (i.e. time-lapse photos, daily routines, video perspectives, object portfolios, and event reports). According to researchers who used this technique, it helps understand nonhumans' perspective and world by revealing rich insights into nonhumans' inner life, their relations with other objects and people, and their subjectivities, and opening a way for empathy between humans and nonhumans. It also allows for an imaginative conversation between humans and nonhumans (who are not able to speak for themselves) through performance. However, they highlight that while the entire process may seem to revolve around the interviewee, both the interviewee's performance and the interviewer's questioning technique are important. The interviewer should ask questions as if speaking to a nonhuman, ensuring an understanding of nonhuman perspectives.

*Umwelt-sketch* (De Roo & Ganzevles, 2023) aims to discover how different nonhumans perceive and interact with their environment. The method is based on the concept of Umwelt, which argues that organisms interact uniquely with the world, guided by their sensory abilities and behaviors. With this method, designers are encouraged to think from the perspective of nonhumans and try to visualize the world from their perspective. According to the authors, this method helps understand nonhuman perspectives and worlds by encouraging sensitivity towards nonhuman actors, exploring different nonhumans on the site, and empathically connecting with them. Furthermore, the method allows participants to decenter human perspectives and oscillate between human and nonhuman perspectives. The authors highlight the importance of gathering information about nonhumans through desk research and experts. They point out that although the method can be successful in understanding nonhuman perspectives, these perspectives may not be adequately represented in the final designs.

Different from other methods, the *Animal Diplomacy Bureau's bird games* (Phillips & Kau, 2019) is a product that aims to immerse participants in the experience of thinking and acting like nonhumans (birds) by taking the role of birds and aiming to gather food points. The game seamlessly integrates elements of the physical with a semi-fictive world. It is constructed upon real life, mirroring the complex dynamics observed among various bird species within a park setting. Moreover, the game's rules are designed to incorporate authentic bird behaviors and interactions. According to the authors, this game helps understand nonhuman perspectives by making participants think about how nonhumans experience parks, encourages discussions about urban life and nature, and enables them to connect with the nonhuman world.

### Reflection

Methods designed to foster reflection primarily involve spending time with nonhuman stakeholders and engaging in self-reflection through these experiences. These include *noticing* and *breathing-with*.

*Noticing* is utilized by three different studies in the literature: with birds (Biggs et al., 2021), with knots (Oogjes & Wakkary, 2022), and with plants and microbes (Rosén et al., 2022). Authors use *noticing* as a reflection method with mainly similar bases. They recommend maintaining openness to new experiences when utilizing this method. Furthermore, they suggest that in addition to experiencing these encounters with an agenda and purpose, the nonhuman entity with whom the experience is shared should also contribute to shaping these experiences. They note that certain interactions may originate from the initiation of the nonhuman entity.

On the other hand, the scope varies slightly in their research. While Biggs et al. primarily focus on establishing intimacy with the nonhuman world through making, Oogjes and Wakkary emphasize that *noticing* can be achieved through embracing heterogeneity and tension rather than seeking harmony or unity. Rosén et al., on the other hand, argue that what is noticed is not coincidental; rather, it is influenced by factors such as cultural background, values, and beliefs. Additionally, they suggest that *noticing* is not merely about observing the surroundings but understanding systematic patterns and "recognizing oneself as part of these interdependencies".

*Noticing* helps in understanding the nonhuman world by bringing attention to the presence of various nonhumans in the environment and revealing previously unnoticed relationships among them (Oogjes & Wakkary, 2022; Rosén et al., 2022). It also helps in understanding the perspectives of nonhuman entities (Rosén et al., 2022) and facilitates a deeper connection with nature (Biggs et al., 2021; Rosén et al., 2022). Biggs et al. add that *noticing* enables this connection through different sensory experiences and helps uncover "new 'edges' of the problem".

All authors using *noticing* emphasize the importance of systematically documenting observations during the noticing process. Rosén et al. highlight the potential benefits of utilizing technology to experience different sensory encounters. Finally, Biggs et al. describe the noticing process as "a psychological labor," noting the challenges faced in adapting oneself to MTH thinking as a designer and researcher.

*Breathing-with* is a design method that broadens the concept of the body in design processes to encompass nonhumans (Fritsch et al., 2023). It involves using breathing to foster an expanded bodily awareness, including material, living, and social aspects of the nonhumans. It highlights intimacy, and vulnerability, and navigates the interconnected relationships between humans and nonhumans. The authors used *breathing-with* as an exploration method with inanimate natural (air) and things (wings) in two different ways. First, Breathing Commons involves a series of sessions and workshops exploring breathing as an individual and a collective experience. Breathing Wings, secondly, is grounded in an interactive wearable prototype shaped like wings, to intimately explore the relationship between the human body and its material extension.

*Breathing-with* facilitates the recognition of the interconnection between humans and nonhumans (such as human-air and human-material (wings)), acknowledging the presence of nonhuman agencies (such as the liveliness of the latex). It helps in understanding the human world by observing various nonhuman stakeholders and exploring ways to coexist with the nonhuman world.

## Discussion

This paper presents an analysis of exploration and synthesis methods used in design research papers exploring nonhuman perspectives. This analysis revealed a shortage of synthesis methods versus exploration methods, notably prioritizing animals and things over plants and microbes, and resulted in the identification of the strengths and weaknesses of various methods. We discuss two key points considering these findings.

### From Unidimensional User Representations to Entangled Representations

With the introduction of MTH Design, sustainable design studies evolve beyond prioritizing efficiency, perceiving nature merely as a resource and viewing humans as separate from nature (Light et al., 2017; Rupprecht et al., 2020). Discussions surrounding MTH highlight the entanglements and interdependencies among humans, urban environments, and the natural world (Haraway, 2013; Moore, 2017). Therefore, sustainability and envisioning sustainable futures require a holistic perspective that goes beyond individual human or nonhuman perspectives. MTH Sustainable Design advocates for considering these stakeholders as integral parts of a larger interconnected system, fostering a more comprehensive understanding of sustainability, and facilitating the development of inclusive designs for the future. This notion indicates a need for tools that can enable designers and researchers to collect data about nonhumans, humans, and their relations, and represent this data in the design process.

In this study, we found that although there are plenty of exploration methods in MTH Design literature, the variety in synthesis methods is limited and they heavily rely on alterations of persona technique as can be seen in Figure 1. The persona technique is widely used in both human-centered and MTH fields. We recognize the practicality of persona methods in terms of providing a representation platform for nonhuman perspectives as a start. The findings also showed a shift in persona usage from representing one nonhuman to multiple nonhumans already. However, the persona has recently been criticized in the literature for being one-dimensional and unable to represent temporal aspects (Roman, 2019). This limitation is particularly evident in MTH contexts, where the user is defined through entanglements. Hence, we identify a need for tools that focus not only on nonhumans individually (looking for motivations, needs, wants, etc.) but also on their relationships with other stakeholders, how they are defined within these relationships, and the dynamics of these relationships. In this respect, we invite researchers and practitioners working in MTH Design to experiment with methods like stakeholder maps, actor-network diagrams, and eco-system loops.

### Need for Understanding Plants and Microbes

This study revealed the scarcity of methods proposed for plants and microbiomes in comparison to animals and things. While 23 studies focus on animals and things, there are only 5 studies focused on plants, 6 on microbes, and 4 on inanimate natural phenomena. One reason for this could be the increasing popularity of IoT devices and research, coupled with the fact that Animal-Computer Interaction (ACI) has been studied since the 1990s (Mancini, 2011). Additionally, plant-computer interaction (PCI) research and studies with microbes are more recent compared to studies focusing on animals, indicating that these studies represent “the margin of the margin” (Aspling et al., 2016).

Based on the knowledge acquired from our overview and literature, we believe that there is a need for an increase in studies related to plants in the MTH field. Plants maintain the soil structure, microbial communities, and ecosystem functions. They support a greater diversity of habitats for microbial communities and prevent soil disturbance, all of which are very important for biodiversity and sustainable environments (Vezzani et al., 2018). They are vital for life on Earth, yet they face significant extinction threats due to "plant blindness", a concept referring to the lack of attention toward plants, where their importance in nature and human life is often overlooked (Thomas et al., 2022). Microbes are used as biofertilizers, biopesticides, and bioinsecticides in sustainable agriculture, which are important for promoting soil health and protecting crops. They are also used for fermentation and recycling waste materials are beneficial for ensuring high protein content (Kalsoom et al., 2020). Thus, considering the growing interest in sustainability, future challenges such as food scarcity (United Nations, 2023), plants’ crucial role, and possible design spaces for microbiomes in a sustainable future, we encourage design researchers and practitioners to plan and execute studies that involve plants and microbes, as well as developing new methods that could help understand these species.

## Conclusion

In this paper, we examined the methods used in early-stage design processes to integrate nonhuman perspectives, particularly focusing on MTH Design. We reviewed 30 papers to analyze the exploration and synthesis techniques employed, noting a scarcity of synthesis methods compared to exploration methods. Additionally, we categorized the methods based on their objectives and discussed their strengths and weaknesses according to the authors’ reflections, emphasizing the need for more inclusive and comprehensive design methods to address the challenges of sustainability and climate change. While the literature predominantly focuses on animals and things, there's a pressing need to develop methods that encompass plants and microbes. Furthermore, the MTH Design field needs more synthesis methods focusing on the entanglements between different stakeholders.

## Acknowledgments

This work was supported by the BAGEP Award of the Science Academy.

## References

Aspling, F., Wang, J., & Juhlin, O. (2016). Plant-computer interaction, beauty and dissemination. *Proceedings of the Third International Conference on Animal-Computer Interaction*, 1–10. https://doi.org/10.1145/2995257.2995393

Bain, I., & Tomitsch, M. (2022). Designing for Personas That Don't Have a Voice: Reflections on Designing a Mobile Application for Collecting Biodiversity Data. *ACM International Conference Proceeding Series*, 229–234. https://doi.org/10.1145/3572921.3572945

Behzad, A., Wakkary, R., Oogjes, D., Zhong, C., & Lin, H. (2022). Iterating through Feeling-with Nonhuman Things. *CHI Conference on Human Factors in Computing Systems Extended Abstracts*, 1–6. https://doi.org/10.1145/3491101.3519860

Benöhr, J., Brinksma, M., Donihue, R., Faro, D., Lara, A., Virik, K. L., León, A. P. de, Toro, C., Gygli, B., Romo, D., & Walther, F. E. (2022). Ecopolitical Mapping: A Multispecies Research Methodology for Environmental Communication. *CS*, *36*, 317–343. https://doi.org/10.18046/RECS.I36.5275

Biggs, H. R., Bardzell, J., & Bardzell, S. (2021). Watching Myself Watching Birds. *Proceedings of the 2021 CHI Conference on Human Factors in Computing Systems*, 1–16. https://doi.org/10.1145/3411764.3445329

Biggs, H. R., & Desjardins, A. (2020). High Water Pants: Designing Embodied Environmental Speculation. *Proceedings of the 2020 CHI Conference on Human Factors in Computing Systems*, 1–13. https://doi.org/10.1145/3313831.3376429

Braidotti, R. (2013). *The Posthuman*. www.politybooks.com

Ceschin, F., & Gaziulusoy, I. (2016). Evolution of design for sustainability: From product design to design for system innovations and transitions. *Design Studies*, *47*, 118–163. https://doi.org/10.1016/j.destud.2016.09.002

Chang, W.-W., Giaccardi, E., Chen, L.-L., & Liang, R.-H. (2017). "Interview with Things." *Proceedings of the 2017 Conference on Designing Interactive Systems*, 1001–1012. https://doi.org/10.1145/3064663.3064717

Chen, D., Seong, Y. A., Ogura, H., Mitani, Y., Sekiya, N., & Moriya, K. (2021). Nukabot: Design of Care for Human-Microbe Relationships. *Conference on Human Factors in Computing Systems - Proceedings*. https://doi.org/10.1145/3411763.3451605

Cho, H., Lee, J., Ku, B., Jeong, Y., Yadgarova, S., & Nam, T.-J. (2023). ARECA: A Design Speculation on Everyday Products Having Minds. *Proceedings of the 2023 ACM Designing Interactive Systems Conference*, 31–44. https://doi.org/10.1145/3563657.3596002

Cila, N., Caldwell, M., Tynan-O'mahony, F., Speed, C., & Rubens, N. (2015). *LISTENING TO AN EVERYDAY KETTLE: HOW CAN THE DATA OBJECTS COLLECT BE USEFUL FOR DESIGN RESEARCH?* http://sites.thehagueuniversity.com/pinc2015/home

Cila, N., & Giaccardi, E. (2015). *Thing-centered narratives: A study of object personas OxChain View project Resourceful Ageing View project*. http://www.moma.org/interactives/exhibitions/2001/workspheres

Clarke, R., Heitlinger, S., Foth, M., Disalvo, C., Light, A., & Forlano, L. (2018). More-than-Human Urban Futures: Speculative Participatory Design to Avoid Ecocidal Smart Cities. *Proceedings of the 15th Participatory Design Conference: Short Papers, Situated Actions, Workshops and Tutorial - Volume 2*, *18*. https://doi.org/10.1145/3210604

Clarke, R., Heitlinger, S., Light, A., Forlano, L., Foth, M., & DiSalvo, C. (2019). More-than-human participation: Design for sustainable smart city futures. *Interactions*, *26*(3), 60–63. https://doi.org/10.1145/3319075

Coskun, A., Cila, N., Nicenboim, I., Frauenberger, C., Wakkary, R., Hassenzahl, M., Mancini, C., Giaccardi, E., & Forlano, L. (2022). More-than-human Concepts, Methodologies, and Practices in HCI. *Conference on Human Factors in Computing Systems - Proceedings*. https://doi.org/10.1145/3491101.3516503

Coulton, P., & Lindley, J. G. (2019). More-Than Human Centred Design: Considering Other Things. *Design Journal*, *22*(4), 463–481. https://doi.org/10.1080/14606925.2019.1614320

Cox, E., Mancini, C., & Ruge, L. (2020). Understanding Dogs’ Engagement with Interactive Games. *Proceedings of the Seventh International Conference on Animal-Computer Interaction*, 1–12. https://doi.org/10.1145/3446002.3446122

De Roo, B., & Ganzevles, G. A. (2023). The Umwelt-sketch as More-than-human Design Methodology Decentering the design process from human-centered towards more-than-human-centered. *DIS 2023 Companion: Companion Publication of the 2023 ACM Designing Interactive Systems Conference*, 203–206. https://doi.org/10.1145/3563703.3596628

Forkosh, O. (2021). Animal behavior and animal personality from a non-human perspective: Getting help from the machine. *Patterns*, *2*(3), 100194. https://doi.org/10.1016/j.patter.2020.100194

Forlano, L. (2016). Decentering the Human in the Design of Collaborative Cities. *Design Issues*, *32*(3), 42–54. https://doi.org/10.1162/DESI_a_00398

Forlano, L. (2017). Posthumanism and Design. *She Ji: The Journal of Design, Economics, and Innovation*, *3*(1), 16–29. https://doi.org/10.1016/J.SHEJI.2017.08.001

Frawley, J. K., Dyson, L. E., Frawley, J., & Dyson, L. E. (2014). Animal personas: Acknowledging non-human stakeholders in designing for sustainable food systems. *Proceedings of the 26th Australian Computer-Human Interaction Conference, OzCHI 2014*, 21–30. https://doi.org/10.1145/2686612.2686617

French, F., Mancini, C., & Sharp, H. (2020). More than human aesthetics: Interactive enrichment for elephants. *DIS 2020 - Proceedings of the 2020 ACM Designing*

*Interactive Systems Conference*, 1661–1672. https://doi.org/10.1145/3357236.3395445

Fritsch, J., Tsaknaki, V., Ryding, K., & Hasse Jørgensen, S. (2023). 'Breathing-with': a design tactic for the more-than-human. *Human-Computer Interaction*. https://doi.org/10.1080/07370024.2023.2275760

Giaccardi, E., Cila, N., Speed, C., & Caldwell, M. (2016). Thing Ethnography. *Proceedings of the 2016 ACM Conference on Designing Interactive Systems*, 377–387. https://doi.org/10.1145/2901790.2901905

Giaccardi, E., Speed, C., Cila, N., & Caldwell, M. L. (2020). Things as Co-Ethnographers: Implications of a Thing Perspective for Design and Anthropology. In *Design Anthropological Futures* (pp. 235–248). Routledge. https://doi.org/10.4324/9781003085188-19

Häggström, M. (2019). Lived experiences of being-in-the-forest as experiential sharing with the more-than-human world. *Environmental Education Research*, *25*(9), 1334–1346. https://doi.org/10.1080/13504622.2019.1633275

Haraway, D. (2013). *When Species Meet*. U of Minnesota Press.

Harman, G. (2017). *Object orientated ontology: A new theory*. 1–20, 253–263. https://www.penguin.co.uk/books/295720/object-oriented-ontology-by-harman-graham/9780241269152

Harrison, M. A., & Hall, A. E. (2010). Anthropomorphism, empathy, and perceived communicative ability vary with phylogenetic relatedness to humans. *Journal of Social, Evolutionary, and Cultural Psychology*, *4*(1), 34–48. https://doi.org/10.1037/H0099303

Hastrup, K. (2015). *Anthropology and Nature* (K. Hastrup, Ed.; 1st ed.). Routledge.

Heitlinger, S., Houston, L., Taylor, A., & Catlow, R. (2021). Algorithmic Food Justice: Co-Designing More-than-Human Blockchain Futures for the Food Commons. *Proceedings of the 2021 CHI Conference on Human Factors in Computing Systems*, 1–17. https://doi.org/10.1145/3411764.3445655

Hirskyj-Douglas, I., Read, J. C., & Horton, M. (2017). Animal Personas: Representing Dog Stakeholders in Interaction Design. *HCI 2017: Digital Make Believe - Proceedings of the 31st International BCS Human Computer Interaction Conference, HCI 2017*, *2017-July*. https://doi.org/10.14236/EWIC/HCI2017.37

Kalsoom, M., Rehman, U. R., Shafique, T., Junaid, S., Khalid, N., Adnan, M., Zafar, I., Tariq, M. A., Raza, M. A., Zahra, A., & Ali[5], H. (2020). *BIOLOGICAL IMPORTANCE OF MICROBES IN AGRICULTURE, FOOD AND PHARMACEUTICAL INDUSTRY: A REVIEW*. *8*, 2020.

Kirk, R. G. W., Pemberton, N., & Quick, T. (2019). Being well together? Promoting health and well-being through more than human collaboration and companionship. *Medical Humanities*, *45*(1), 75. https://doi.org/10.1136/MEDHUM-2018-011601

Lacetera, N. (2019). Impact of climate change on animal health and welfare. *Animal Frontiers*, *9*(1), 26–31. https://doi.org/10.1093/af/vfy030

Latour, B. (1996). *On actor-network theory*. Soziale Welt. https://www.jstor.org/stable/40878163

Liedtka Jeanne, & Ogilvie Tim. (2011). *Designing for Growth: A Design Thinking Toolkit for Managers*. Columbia University Press.

Light, A., Shklovski, I., & Powell, A. (2017). Design for Existential Crisis. *CHI* , 722–734. https://doi.org/10.1145/3027063.3052760

Liu, J., Byrne, D., & Devendorf, L. (2018). Design for Collaborative Survival. *Proceedings of the 2018 CHI Conference on Human Factors in Computing Systems*, 1–13. https://doi.org/10.1145/3173574.3173614

Liu, S.-Y. (Cyn), Bardzell, J., & Bardzell, S. (2018). Photography as a Design Research Tool into Natureculture. *Proceedings of the 2018 Designing Interactive Systems Conference*, 777–789. https://doi.org/10.1145/3196709.3196819

Livio, M., & Devendorf, L. (2022). The Eco-Technical Interface: Attuning to the Instrumental. *Conference on Human Factors in Computing Systems - Proceedings*. https://doi.org/10.1145/3491102.3501851

Macruz, A., Daneluzzo, M., & Tawakul, H. (2023). *Performative Ornament: Enhancing Humidity and Light Levels for Plants in Multispecies Design*. 478–487. https://doi.org/10.1007/978-981-19-8637-6_41

Malhi, Y., Franklin, J., Seddon, N., Solan, M., Turner, M. G., Field, C. B., & Knowlton, N. (2020). Climate change and ecosystems: threats, opportunities and solutions. *Philosophical Transactions of the Royal Society B: Biological Sciences*, *375*(1794), 20190104. https://doi.org/10.1098/rstb.2019.0104

Mancini, C. (2011). Animal-Computer Interaction: A Manifesto. *Interactions*, *18*(4), 69–73.

Markard, J., Raven, R., & Truffer, B. (2012). Sustainability transitions: An emerging field of research and its prospects. *Research Policy*, *41*(6), 955–967. https://doi.org/10.1016/J.RESPOL.2012.02.013

Metzger, J. (2014). Spatial planning and/as caring for more-than-human place. In *Environment and Planning A* (Vol. 46, Issue 5, pp. 1001–1011). Pion Limited. https://doi.org/10.1068/a140086c

Metzger, J. (2019). A more-than-human approach to environmental planning. *The Routledge Companion to Environmental Planning*, 190–199. https://doi.org/10.4324/9781315179780-20/HUMAN-APPROACH-ENVIRONMENTAL-PLANNING-JONATHAN-METZGER

Metzger, J., & Lindblad, J. (2020). Dilemmas of Sustainable Urban Development. In *Dilemmas of Sustainable Urban Development*. Routledge.

https://doi.org/10.4324/9780429294457/DILEMMAS-SUSTAINABLE-URBAN-DEVELOPMENT-JONATHAN-METZGER-JENNY-LINDBLAD

Moore, J. W. (2017). The Capitalocene, Part I: on the nature and origins of our ecological crisis. *The Journal of Peasant Studies*, *44*(3), 594–630. https://doi.org/10.1080/03066150.2016.1235036

Murray-Rust Katerina Gorkovenko DMurray-Rust, D., KGorkovenko, edacuk, Burnett Daniel Richards DBurnett, D., DRichards, lancasteracuk, Murray-Rust, D., Gorkovenko, K., Burnett, D., & Richards, D. (2019). Entangled ethnography: Towards a collective future understanding. *ACM International Conference Proceeding Series*. https://doi.org/10.1145/3363384.3363405

Oogjes, D., & Wakkary, R. (2022). Weaving Stories: Toward Repertoires for Designing Things. *CHI Conference on Human Factors in Computing Systems*, 1–21. https://doi.org/10.1145/3491102.3501901

Pettersen, I. N., Geirbo, H. C., & Johnsrud, H. (2018). The tree as method: Co-creating with urban ecosystems. *ACM International Conference Proceeding Series*, *2*. https://doi.org/10.1145/3210604.3210653

Phillips, R., & Kau, K. (2019). Gaming for Active Nature Engagement. Animal Diplomacy Bureau: designing games to engage and create player agency in urban nature. *The Design Journal*, *22*(sup1), 1587–1602. https://doi.org/10.1080/14606925.2019.1594993

*PlantWave*. (2020). https://plantwave.com/en-tr

Pollastri, S., Griffiths, R., Dunn, N., Cureton, P., Boyko, C., Blaney, A., & De Bezenac, E. (2021). More-Than-Human Future Cities: From the design of nature to designing for and through nature. *Media Architecture Biennale 20*, 23–30. https://doi.org/10.1145/3469410.3469413

Porter, N. (2019). Training Dogs to Feel Good: Embodying Well-being in Multispecies Relations. *Medical Anthropology Quarterly*, *33*(1), 101–119. https://doi.org/10.1111/MAQ.12459

Portocarrero, E., Dublon, G., Paradiso, J., & Bove, M. V. (2015). ListenTree: Audio-haptic display in the natural environment. *Conference on Human Factors in Computing Systems - Proceedings*, *18*, 395–398. https://doi.org/10.1145/2702613.2725437

Prost, S., Pavlovskaya, I., Meziant, K., Vlachokyriakos, V., & Crivellaro, C. (2021). Contact Zones. *Proceedings of the ACM on Human-Computer Interaction*, *5*(CSCW1), 1–24. https://doi.org/10.1145/3449121

Reddy, A., Kocaballi, A. B., Nicenboim, I., Søndergaard, M. L. J., Lupetti, M. L., Key, C., Speed, C., Lockton, D., Giaccardi, E., Grommé, F., Robbins, H., Primlani, N., Yurman, P., Sumartojo, S., Phan, T., Bedö, V., & Strengers, Y. (2021). Making Everyday Things Talk: Speculative Conversations into the Future of Voice Interfaces at Home.

*Extended Abstracts of the 2021 CHI Conference on Human Factors in Computing Systems*, 1–16. https://doi.org/10.1145/3411763.3450390

Reddy, A., Nicenboim, I., Pierce, J., & Giaccardi, E. (2021). Encountering ethics through design: a workshop with nonhuman participants. *AI and Society*, *36*(3), 853–861. https://doi.org/10.1007/S00146-020-01088-7/FIGURES/4

Roman, C. (2019, February 27). *The Problem with Personas*. https://medium.com/typecode/the-problem-with-personas-b6734a08d37a

Rosén, A. P., Normark, M., & Wiberg, M. (2022). *Towards More-Than-Human-Centred Design: Learning from Gardening*. International Journal of Design. http://www.ijdesign.org/index.php/IJDesign/article/view/4402

Rupprecht, C. D. D., Vervoort, J., Berthelsen, C., Mangnus, A., Osborne, N., Thompson, K., Urushima, A. Y. F., Kóvskaya, M., Spiegelberg, M., Cristiano, S., Springett, J., Marschütz, B., Flies, E. J., McGreevy, S. R., Droz, L., Breed, M. F., Gan, J., Shinkai, R., & Kawai, A. (2020). Multispecies sustainability. In *Global Sustainability*. Cambridge University Press. https://doi.org/10.1017/sus.2020.28

*Scencosme – Phonopholium*. (2018). https://wakeupscreaming.com/gregory-lasserre-anais-met-den-ancxt-scenocosme/

Scuri, S., Ferreira Nuno, M., Nunes, J., Nisi, V., Ferreira, M., Mulligan, C. 2022, & Chi ', (. (2022). Hitting the Triple Bottom Line. *CHI '22: Proceedings of the 2022 CHI Conference on Human Factors in Computing Systems*, 1–19. https://doi.org/10.1145/3491102.3517518

Sheikh, H., Gonsalves, K., & Foth, M. (2021). Plant(e)tecture Towards a Multispecies Media Architecture Framework for amplifying Plant Agencies. *MAB20: Media Architecture Biennale 20*. https://doi.org/10.1145/3469410.3469419

Smith, N., Bardzell, S., & Bardzell, J. (2017). Designing for cohabitation: Naturecultures, hybrids, and decentering the human in design. *Conference on Human Factors in Computing Systems - Proceedings*, *2017-May*, 1714–1725. https://doi.org/10.1145/3025453.3025948

Søndergaard, M. L. J. (2023). What mosses can teach us about design fabulations and feminist more-than-human care. *Human–Computer Interaction*. https://doi.org/10.1080/07370024.2023.2269893

Steiner, H., Johns, P., Roseway, A., Quirk, C., Gupta, S., & Lester, J. (2017). Project Florence. *Proceedings of the 2017 CHI Conference Extended Abstracts on Human Factors in Computing Systems*, 1415–1420. https://doi.org/10.1145/3027063.3052550

Talgorn, E., & Ullerup, H. (2023). Invoking 'Empathy for the Planet' through Participatory Ecological Storytelling: From Human-Centered to Planet-Centered Design. *Sustainability 2023, Vol. 15, Page 7794*, *15*(10), 7794. https://doi.org/10.3390/SU15107794

Thomas, H., Ougham, H., & Sanders, D. (2022). Plant blindness and sustainability. *International Journal of Sustainability in Higher Education*, *23*(1), 41–57. https://doi.org/10.1108/IJSHE-09-2020-0335

Tironi, M., Chilet, M., Marín, C. U., & Hermansen, P. (2023). Design for More-Than-Human Futures: Towards Post-Anthropocentric Worlding. *Design for More-Than-Human Futures: Towards Post-Anthropocentric Worlding*, 1–172. https://doi.org/10.4324/9781003319689

Tomitsch, M., Fredericks, J., Vo, D., Frawley, J., & Foth, M. (2021). Non-human Personas. Including Nature in the Participatory Design of Smart Cities. *Interaction Design and Architecture(s)*, *50*, 102–130. https://doi.org/10.55612/s-5002-050-006

Turner, J., & Morrison, A. (2020). Designing Slow Cities for More Than Human Enrichment: Dog Tales—Using Narrative Methods to Understand Co-Performative Place-Making. *Multimodal Technologies and Interaction 2021, Vol. 5, Page 1*, *5*(1), 1. https://doi.org/10.3390/MTI5010001

United Nations. (2023). *The Sustainable Development Goals Report 2023: Special Edition*. https://unstats.un.org/sdgs/report/2023/The-Sustainable-Development-Goals-Report-2023.pdf

Vezzani, F. M., Anderson, C., Meenken, E., Gillespie, R., Peterson, M., & Beare, M. H. (2018). The importance of plants to development and maintenance of soil structure, microbial communities and ecosystem functions. *Soil and Tillage Research*, *175*, 139–149. https://doi.org/10.1016/j.still.2017.09.002

Weheliye, A. G. (2014). *Habeas viscus : racializing assemblages, biopolitics, and black feminist theories of the human*.

Westerlaken, M., & Gualeni, S. (2016). Becoming with. *Proceedings of the Third International Conference on Animal-Computer Interaction*, 1–10. https://doi.org/10.1145/2995257.2995392

Wolff, A., Knutas, A., Pässilä, A., Lautala, J., Kantola, L., & Vainio, T. (2021, May 8). Designing SciberPunks as Future Personas for More than Human Design. *Conference on Human Factors in Computing Systems - Proceedings*. https://doi.org/10.1145/3411763.3443443

Wolniak, R. (2017). The Design Thinking method and its stages. *Systemy Wspomagania w Inżynierii Produkcji*, *Vol. 6, iss. 6*.